\documentclass[%
 aip,
 amsmath,amssymb,
 reprint,%
]{revtex4-1}

\usepackage[usenames,dvipsnames,svgnames,table]{xcolor}
\usepackage{graphicx}% Include figure files
\usepackage{dcolumn}% Align table columns on decimal point
\usepackage{bm}% bold math
\usepackage[normalem]{ulem}
\usepackage{soul}
\usepackage[utf8]{inputenc}
\usepackage[T1]{fontenc}
\usepackage{mathptmx}
\usepackage{etoolbox}
\usepackage{verbatim}

\newcommand{\blue}[1]{\textcolor{black}{#1}}
\newcommand{\mage}[1]{\textcolor{black}{#1}}
\newcommand{\red}[1]{\textcolor{black}{#1}}

\newcommand{\av}[1]{\langle {#1} \rangle}
\newcommand{\stk}[1]{\ifmmode\text{\sout{\ensuremath{#1}}}\else\sout{#1}\fi}

\makeatletter
\def\@email#1#2{%
 \endgroup
 \patchcmd{\titleblock@produce}
  {\frontmatter@RRAPformat}
  {\frontmatter@RRAPformat{\produce@RRAP{*#1\href{mailto:#2}{#2}}}\frontmatter@RRAPformat}
  {}{}
}%
\makeatother
\begin{document}

\preprint{AIP/123-QED}

\title[]{\red{Chimera-like} states in neural networks and power systems}
% Force line breaks with \\
\author{Shengfeng Deng}%
\affiliation{School of Physics and Information Technology, Shaanxi Normal University, Xi’an 710062, China}%
\author{G\'eza \'Odor$^{*}$}
\email{odor.geza@ek-cer.hu}
 \affiliation{Institute of Technical Physics and Materials Science, \red{HUN-REN} Centre for Energy Research, P.O. Box 49, H-1525 Budapest, Hungary}
\date{\today}% It is always \today, today,

\begin{abstract}
 Partial, frustrated synchronization and \red{chimera-like} states are expected to occur in Kuramoto-like models
 if the spectral dimension of the underlying graph is low: $d_s < 4$. We provide numerical evidence that 
 this really happens in case of the high-voltage power grid of Europe ($d_s < 2$), \blue{a large human connectome (KKI113)} and in case of the largest, exactly known brain network corresponding to the fruit-fly (FF) connectome ($d_s < 4$), even though their 
 graph dimensions are much higher, i.e.: $d^{EU}_g\simeq 2.6(1)$ and $d^{FF}_g\simeq 5.4(1)$, $d^{\mathrm{KKI113}}_g\simeq 3.4(1)$.
 We provide local synchronization results of the first- and second-order (Shinomoto) Kuramoto models by numerical solutions on the FF and the European power-grid graphs, respectively, and show the emergence of \red{chimera-like} patterns on the graph community level as well as by the local order parameters.
\end{abstract}

\maketitle

\begin{quotation}
We show that Kuramoto oscillator models on large neural connectome graph of the fruit-fly, 
a human brain, as well as on the power-grid of Europe produce \red{chimera-like} states.
This is in agreement with the low spectral dimensions that we calculated by the eigenvalue spectra of the Laplacian of these networks. We compare these results with the topological dimension measurements and previous simulations, strengthening that frustrated synchronization should occur, which can generate slow relaxations, obtained in previous studies within the neighborhood of the synchronization 
transition point.
\end{quotation}

\section{Introduction \label{sec:1}}

Synchronization phenomena are very widespread in nature and the understanding of their behavior is in the focus
of interest. In neural systems, like the brain oscillatory behavior of building elements has been measured
by different techniques, while in case of power grids, the alternating currents can also be described by
coupled oscillators. Both systems are expected to operate close to the synchronization transition point.
In case of the normal brain, self-tuning to the critical point is hypothesed~\cite{ChialvBak1999_LearningMistakes}
and confirmed by experiments~\cite{BP03} and theoretical considerations~\cite{MArep}. 
The advantage of criticality is the optimal computational performance, sensitivity as well as dynamically generated long-range memory and interactions~\cite{Larr}.
In case of power grids the competition of supply and demands tune the system close to the synchronization 
transition point~\cite{Car}.

Synchronization models described by the first Kuramoto equation~\cite{kura} have recently 
been investigated on complex networks and partial synchronization was found if the spectral dimension is below $4$ even if generalizations of the Euclidean dimension, the graph and the Hausdorff 
dimension are high or diverge~\cite{millan2019}. 
Partial synchronization is more probable in strongly connected modules
or communities, which also happens both in biological and technical structures. Modular and most often hierarchical 
organization is known in general brain networks~\cite{sporns2010networks}, among others in case of the fruit-fly (FF) connectome~\cite{FlybrainAtl,Flycikk},  as well as in power grids~\cite{POWcikk}. 
Thus synchronization occurs in the strongly coupled modules first, while in the loosely coupled parts, 
nodes may remain desynchronized for the same conditions, which was called frustrated
synchronization~\cite{Villegas_2014}, reminiscent of the semi-critical Griffiths Phases (GP) of condensed 
matters~\cite{Griffiths}. Besides, in these phenomena, fluctuations of the global order parameters diverge in an extended control parameter space. 
Recently this was shown in case of Kuramoto models on brain connectomes~\cite{Villegas_2014,KurCC,KKI18deco,Flycikk,sniccikk} 
as well as in case of power-grids~\cite{POWcikk,Powfailcikk,USAEUPowcikk}.

One can also relate such structures, emerging in these heterogeneous systems, to \red{chimera-like} states, in which subsets of an ensemble of identical, interacting oscillators exhibit distinct dynamical states, such as one group of synchronized oscillators and one group of desynchronized oscillators~\cite{Abrams-Strogatz-2004}.
Firstly chimeras were defined in systems of identical oscillators~\cite{Abrams-Strogatz-2004,PhysRevLett.101.084103}.
In such a case, a non-zero phase lag term is essential for partial synchronization to occur. Realistic models, however, require oscillators to be heterogeneous and \red{chimera-like states have been reported} on \red{complex \mage{networks}~\cite{PhysRevE.89.022914,Scholl2016,Sawicki_2017,zakharova2021chimera}, on human ~\cite{Chouzouris_2018,And2016} as well as on
C-elegans~\cite{Hizanidis2016} neural networks. \mage{Henceforth, by focusing on neural and power-system networks, w}e follow the second line of chimera-like definition \mage{for heterogeneous systems} in our study.}

\blue{
The main criterion to call partially synchronized states as \red{chimera-like} stated in~\cite{Scholl2016} is ``the coexistence of coherent and incoherent domains in space''. The second significant feature of \red{chimera-like} states is the difference of averaged oscillator frequencies. Usually, the oscillators belonging to the coherent domains have identical frequencies, and oscillators from incoherent domains are characterized by higher or lower mean frequencies.}

The purpose of the present study is to provide numerical evidences of such structures by solving Kuramoto equations in seemingly different areas of complex systems, on the largest available brain connectome and on the European high voltage power-grid network. We show 
they are characterized by a low spectral dimension: $d_s<4$.
In these models quenched heterogeneity is present structurally, by the topology of the
graphs as well as by the different self-frequencies of nodes.

\section{Methods \label{sec:2}}
In this section, we detail the models and the methods we applied to describe 
synchronization on different networks.

%%%%%%%%%%%%%%%%%%%%%%%%%%%%%%%%%%%%%%%%%%%%%%%%%%%%%%%%%%%%%%%%%%%%%%%%
\subsection{The first-order Kuramoto model}
%%%%%%%%%%%%%%%%%%%%%%%%%%%%%%%%%%%%%%%%%%%%%%%%%%%%%%%%%%%%%%%%%%%%%%%%%

Several oscillator models have been used in biology, the simplest
possible one is the Hopf model~\cite{Freyer6353}, which has
been used frequently in neuroscience, as it can describe
a critical point with scale-free avalanches, with sharpened
frequency response and enhanced input sensitivity.
The local dynamics of each brain area (node) is described by the normal 
form of a supercritical Hopf bifurcation, also called a 
Landau-Stuart oscillator, which is the canonical
model for studying the transition from noisy to oscillatory dynamics.

Another complex model, describing more non-linearity%
\footnote{ In the weak coupling limit an equivalence with the
integrate-and-fire models~\cite{PolRos15} was shown.}
is the Kuramoto model~\cite{kura,Acebron}, with phases $\theta_i(t)$, 
located at the $N$ nodes of a network, according to the dynamical equation
\begin{equation}
\dot{\theta_i}(t) = \omega_{i}^0 + K \sum_{j} W_{ij}
\sin[ \theta_j(t)- \theta_i(t)]\,.
\label{kureq}
\end{equation}
The global coupling $K$ is the control parameter of this model, by which
we can tune the system between asynchronous and synchronous states. The summation 
is performed over the nearest neighboring nodes, with connections described by the 
\blue{weighted or unweighted} adjacency matrix $W_{ij}$ and $\omega_{i}^0$ denotes the
\blue{quenched} intrinsic frequency of the $i$-th oscillator.
For simplicity, we used the Gaussian distribution with zero mean and unit variance for 
the self-frequency distribution $g(\omega_{i}^0)$ with respect to a rotating frame~\cite{KurCC}.

\blue{Eq. (\ref{kureq}) was originally defined on full graphs, corresponding to mean-field behavior~\cite{chate_prl}. The critical dynamical behavior has been explored on various random graphs~\cite{cmk2016,KurCC}.
Phase transition can happen only above the lower critical dimension $d_l^-=4$ \cite{HPCE}. In lower dimensions, a true, singular phase transition in the $N\to\infty$ limit is not possible, but partial synchronization can emerge with a smooth crossover if the oscillators are
strongly coupled.}
Using this model the resting state critical behavior on large human connectomes, \blue{in particular the so called KKI113}~\cite{KurCC,KKI18deco},  was compared with that of the FF~\cite{Flycikk} on the
global order parameter level and the topology dependence has been pointed out,
which suggested extended fluctuation region and GP-like behavior in case of the human connectomes in contrast with the FF network.

Very recently we have also investigated an extension of Eq.~\eqref{kureq} to the 
Shinomoto--Kuramoto (SK) model, with periodically driven forces~\cite{sakaguchi1988} 
to describe task phase of the brain models~\cite{sniccikk}
\begin{eqnarray}
\label{diffeq}
\dot\theta_j(t) &=& \omega_j^0+K\sum_k W_{jk}\sin[\theta_k(t)-\theta_j(t)] \\
\nonumber
&+& F\sin(\theta_j(t)) + \epsilon\eta_j(t) \ .
\end{eqnarray}
Here $\epsilon$ describes an excitation, with a zero centered, Gaussian random annealed noise $\eta_j(t)$ and a site-dependent periodic
force term, proportional to a coupling $F$, was also added. But in fact, a small $\eta$ proved to be irrelevant, for the synchronization transition, caused by $F$, in the presence of the chaotic noise. 

\blue{
The periodically forced Kuramoto model was studied first by Sakaguchi~\cite{10.1143/PTP.79.39}. However, he found that the state of forced entrainment was not always reached. Macroscopic fractions of the system self-synchronized at a different frequency from that of the drive, indicating that these sub-populations had established their own collective rhythm.
Recently, in~\cite{PhysRevResearch.3.023224} the avalanche behavior
of the SK equation was investigated, using site independent
self-frequencies $\omega_j^0 = \omega$. These authors explored the phase diagram and found a so-called saddle node invariant cycle and a hybrid type of bifurcation besides the forceless Hopf transition. In a very recent publication~\cite{Buend_a_2022} this numerical analysis has been applied for Erd\H{o}s--R\'enyi and hierarchical modular networks, motivated by brain research. Considering quenched
$\omega_j^0$-s, with bi-modal frequency distributions the authors
claim the emergence of Griffiths effects by the broadening of the
synchronization transition region.}

One of the main conclusions of Ref.~\cite{sniccikk} was
that community dependent values of the Hurst exponent $H$ and the $\beta$ 
exponent, measuring self-similarity of time series, varied more with $F>0$, than in the resting state of the brain, corresponding to $F=0$.  Now we shall test this community dependence of $R$ and $\Omega$ in the steady state.

%%%%%%%%%%%%%%%%%%%%%%%%%%%%%%%%%%%%%%%%%%%%%%%%%%%%%%%%%%%%%%%%%%%%%%%%
\subsection{The second-order Kuramoto model}
%%%%%%%%%%%%%%%%%%%%%%%%%%%%%%%%%%%%%%%%%%%%%%%%%%%%%%%%%%%%%%%%%%%%%%%%%

The time evolution of power-grid synchronization is described by the swing
equations~\cite{swing}, set up for mechanical elements (e.g.~rotors in generators 
and motors) with inertia.
It is formally equivalent to the second-order Kuramoto equation~\cite{fila},
for a network of $N$ oscillators with phases $\theta_i(t)$:
\begin{eqnarray}\label{kur2eq}
\dot{\theta_i}(t) & = & \omega_i(t) \\
\dot{\omega_i}(t) & = & \omega_i^0 - \alpha \dot{\theta_i}(t) 
+ K \sum_{j=1}^{N} A_{ij} \sin[ \theta_j(t)- \theta_i(t)] \ . \nonumber
\end{eqnarray}
Here $\alpha$ is the damping parameter, which describes the power dissipation,
or an instantaneous feedback~\cite{Powfailcikk}, $K$ is the global coupling,
related to the maximum transmitted power between nodes; and $A_{ij}$, which
is the adjacency matrix of the network, contains 
admittance elements.
The quenched external drive, denoted by $\omega_i^0$, which is proportional
to the self-frequency of the $i$-th oscillator and carries a dimension of
inverse squared time $[1/s^2]$, describes the power in/out of a given node when
Eq.~(\ref{kur2eq}) is considered to be the swing equation of a coupled AC circuit, 
but here, similar to the first-order Kuramoto model, 
we have chosen it zero centered Gaussian random variable as rescaling invariance of the equation allows to transform it out within a rotating frame. For simplicity, one can assume that $\omega_i(0)$ is drawn
from the same distribution as $\omega_i^0$ and numerically set
$\omega_i(0)=\omega_i^0$, amounting to taking $[s]$=1.
In our present study the following parameter settings were used:
the dissipation factor $\alpha$, is chosen to be equal to $0.4$ to
meet expectations for power grids, with the $[1/s]$ inverse time 
physical dimension assumption.

To characterize the phase transition properties the phase order parameter $R(t)$ 
has been studied for both the first- and second-order Kuramoto models. To ensure the relaxation to the steady states, we measured the Kuramoto phase order parameter
\begin{equation}\label{ordp}
z(t_k) = r(t_k) \exp[i \theta(t_k)] = 1 / N \sum_j \exp{[i \theta_j(t_k)}] \ ,
\end{equation}
where $0 \le r(t_k) \le 1$ gauges the overall coherence and $\theta(t_k)$ 
is the average phase at discrete sampling times $t_k$, which was chosen 
to follow an exponential growth: $t_k = 1 + 1.08^{k}$ to spare memory space. 
The calculation of derivatives was done adaptively at small time steps via the Bulirsch-Stoer stepper~\cite{BS}. 
The sets of equations \eqref{kureq}, \eqref{diffeq} and \eqref{kur2eq} were solved numerically 
for $10^3 - 10^4$ independent initial conditions in ~\cite{USAEUPowcikk},
initialized by different $\omega_i^0$-s and different $\theta_{i}(0)$-s if 
disordered initial phases were invoked. 
Then sample averages for the phases and the frequencies give rise to the Kuramoto order parameter
\begin{equation}\label{KOP}
R(t_k) = \langle r(t_k)\rangle \ ,
\end{equation}
and the variance of the frequencies
\blue{
\begin{equation}\label{FOP}
\Omega(t) = \frac{1}{N} \sum_{j=1}^N (\overline\omega(t)-\omega_j(t))^2  \ .
\end{equation}
}

We don't discuss the time-dependent behavior of the global order parameters 
as this has been investigated in detail in~\cite{USAEUPowcikk,2D3Dcikk}.
In the steady state, which we determined by visual inspection of
$R(t)$ and $\Omega(t)$, we measured their half values and
the standard deviations $\sigma(R(t))$ and $\sigma(\Omega(t))$
in order to locate the transition points. In the paper
we used the $\sigma(R)$, $\sigma(\Omega)$ values, obtained by
sample and time averages in the steady state.

\subsection{Topological and spectral dimensions \label{sec:2A}}

The effective graph (topological) dimension $d_g$ is 
defined by
\begin{equation} \label{topD}
N(r) \sim r^{d_g},
\end{equation}
where we counted the number of nodes $N(r)$ with chemical distance $r$ or less from randomly selected seeds and calculated averages over many trials \cite{shanker2007,deng2015}. In most cases, $d_g$ as obtained from this cluster growing method can be served as an estimation for the more rigorously defined Hausdorff dimension $d_H\approx d_g$ \cite{feder1988}.
In Ref.~\cite{Flycikk} the graph dimension of the FF was estimated to be $d_g^{FF}=5.4(1)$, while in \cite{USAEUPowcikk} we provided a value for the unweighted European power-grid $d_g^{EU}=2.6(1)$.

For regular Euclidean lattices, it has been shown that a true transition to global synchronization under the thermodynamics limit is only possible for $d>4$ in both the first- and the second-order Kuramoto models \cite{hong2005,odor2023}, while for $d\le 4$, there is only a crossover from the asynchronous phase to a partial synchronous phase characterized by an increasingly broadened variance of $R$ and a shifting crossover point $K_c'$ as the system size increases \cite{millan2018,odor2023}. The natural question is then if it is the topological dimension $d_g$ defined in Eq.~\eqref{topD}, or equivalently the Hausdorff dimension $d_H$, that dictates the synchronization properties for general networks which may assume non-integer dimensions. Refs.~\cite{millan2018,millan2019} suggested that the synchronization properties of a general network should be related to the spectral dimension derived from the eigenvalue spectrum of the graph Laplacian matrix, and even more so for the so-called complex network manifolds studied therein, which are constructed out of finite-dimensional simplicies but are characterized by an infinite Hausdorff dimension due to their small-world properties \cite{song2005,millan2018}.

Graph spectral properties of complex networks have been shown to be particularly relevant to the structures of networks~\cite{chung1997}. Following Refs.~\cite{millan2018,millan2019}, we adopt the normalized Laplacian $L$ with elements
\begin{equation}
    L_{ij}=\delta_{ij}-A_{ij}/k_i
    \label{eqs:LnoW}
\end{equation}
for unweighted networks, where $k_i$ denotes the degree of node $i$. Similarly, for weighted networks, the elements of the normalized Laplacian are given by
\begin{equation}
     L_{ij}=\delta_{ij}-W_{ij}/k_i'\,,
    \label{eqs:LWei}
\end{equation}
where $k_i'=\sum_j W_{ji}$ denotes the weighted in-degree of node $i$. The normalized Laplacian has real eigenvalues $0=\lambda_1 \le \lambda_2 \le \dots \le \lambda_N$, the density of which scales as \cite{burioni1996,millan2018}
\begin{equation}
    \rho(\lambda)\simeq \lambda^{d_s/2-1}
    \label{eqs:dsl}
\end{equation}
for $\lambda\ll 1$, where $d_s$ is the spectral dimension. The cumulative density is then given by
\begin{equation}
    \rho_c(\lambda) =\int_0^\lambda d\lambda'\rho(\lambda') \simeq \lambda^{d_s/2}.
    \label{eqs:dsc}
\end{equation}
Since Eqs.~\eqref{eqs:dsl} and \eqref{eqs:dsc} hold for small $\lambda$ values, for the fruit-fly connectome and the European high-voltage power grid network that are going to be studied in Sec.~\ref{sec:3}, with which $N\gg 1$, we will only extract the densities for the first 200 smallest eigenvalues for ease of eigenvalue computation without loss of generality.
\begin{figure}[!tbp]
    \centering
    \includegraphics[width=0.95\columnwidth]{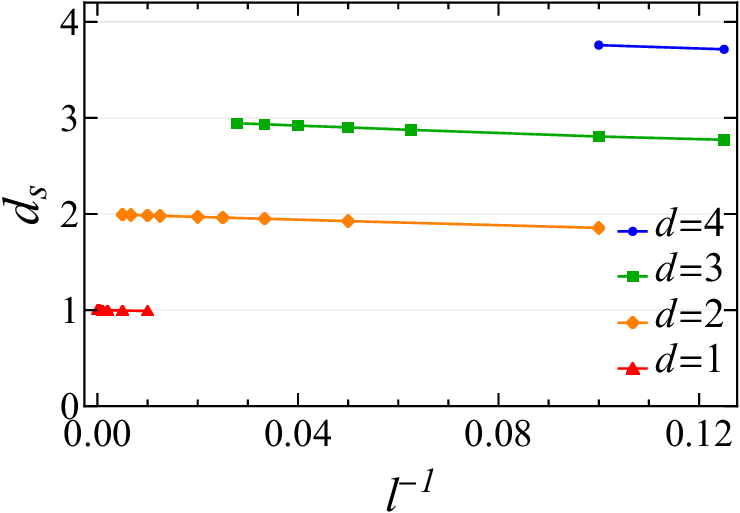}
    \caption{Spectral dimensions of $d$-dimensional regular lattices ($d=1, 2, 3, 4$) of different lateral sizes $l$, measured through Eq.~\eqref{eqs:dsc}. The real dimension of a lattice is asymptotically approached as $l\to \infty$.}
    \label{fig:dslat}
\end{figure}

As illustrated in Fig.~\ref{fig:dslat}, Euclidean lattices in dimension $d$ have spectral dimension $d_s=d$. Therefore in this case the spectral dimension is also equal to the Hausdorff dimension of the lattice, $d_s = d_H$. However, in general, networks can have non-integer spectral dimension $d_s$ not equal to their Hausdorff dimension. Ref.~\cite{millan2018} demonstrated that in lower spectral dimensions $d_s<4$, there is a parameter regime that exhibits frustrated synchronization with spatiotemporal fluctuations even in the stationary state. Then, similar to the emergence of rare regions in Griffiths phases \cite{Griffiths}, one should expect to observe states with rare regions--usually called ``\red{chimera-like} states''--in such frustrated synchronization as well, as we will demonstrate in what follows with the aid of the local order parameter of Kuramoto models.

\begin{figure}[!tbp]
    \centering
    \includegraphics[width=0.98\columnwidth]{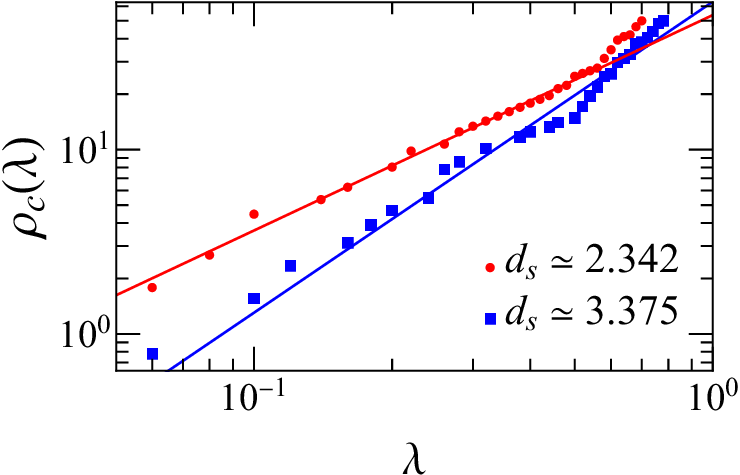}
    \caption{The cumulative eigenvalue densities of the unweighted (blue solid square) and the weighted (red dot) FF network, extracted from the first 200 smallest eigenvalues.}
    \label{fig:dsff}
\end{figure}

\subsection{Local order parameters of the first- and second-order Kuramoto models \label{sec:2B}}

To investigate the heterogeneity further, we measured the local Kuramoto order parameter, defined as the partial sum of phases for the neighbors of node $i$
\begin{equation}\label{LCO}
        r_i(t)= \frac{1}{N_{\mathrm{i.neigh}}}
                \left|\sum_j^{N_{\mathrm{i.neigh}}}  A_{ij} e^{i
        \theta_j(t)}\right|  \ .
\end{equation}
This local Kuramoto measure was firstly suggested by Restrepo
\textit{et al.} \cite{restrepo2005,schroder2017} to quantify the local
synchronization of nodes, which allows us to visualize regions
of synchronized/unsynchronized chimera-like behavior and which will be the main quantity of interest of this paper.

\section{\protect\red{Chimera-like} states \label{sec:3}}

\subsection{\protect\red{Chimera-like} states in the fruit-fly (FF) and in a human connectome} \label{sec:3B}

First, we examine \red{chimera-like} states in the first-order Kuramoto model on the FF connectome. Connectomes are defined as structural networks of neural connections of the brain~\cite{sporns2010networks}. For the fruit fly, we used the hemibrain dataset (v1.0.1)
from~\cite{down-hemibrain1.0.1}, which has $N_{FF}=21\,662$ nodes and $E_{FF}=3\,413\,160$ edges, out of which the largest single connected component contains $N=21\,615$ and
$E=3\,410\,247$ directed and weighted edges, with weights being the number of connections between a pair of nodes. \blue{These parameters come from the database after discarding the unconnected parts.}
The number of incoming edges varies between $1$ and $2708$.
The weights are integer numbers, varying between $1$ and $4299$.
The average node degree is $\langle k\rangle= 315.129$ 
(for the in-degrees it is: $157.6$), while the average weighted degree 
is $\langle w\rangle= 628$.
The adjacency matrix, visualized in~\cite{Flycikk}, shows a weak
hierarchical modular structure, however, it is not random.
For example, the degree distribution is much wider than that of a 
random graph and exhibits a fat tail.
The analysis in~\cite{Flycikk} found a weight distribution $p(W_{ij})$
with a heavy tail, and assuming a power-law (PL) form, a decay exponent $2.9(2)$ 
could be fitted for the $W_{ij} > 100$ region.
The modularity quotient of a network is defined by \cite{Newman2006-bw}
\begin{equation}
Q=\frac{1}{N\av{k}}\sum\limits_{ij}\left(A_{ij}-
\frac{k_ik_j}{N\av{k}}\right)\delta(g_i,g_j)\,.
\end{equation}
The maximum of this value corresponds to the optimal community structure, \blue{and} characterizes how modular a network is, where $\delta(g_i,g_j)$ is $1$ when nodes $i$ and $j$ were found to
be in the same community $g$, or $0$ otherwise.
Community detection algorithms based on modularity optimization get
the closest to the actual modular properties of the network.
The modularity was calculated using community structures detected by the Louvain method \cite{Blondel2008},  from which we obtained
$Q_{FF} \approx 0.631$ \cite{sniccikk}.
The effective graph (topological) dimension, obtained by the
breadth-first search algorithm is $d_g^{FF}=5.4(5)$.

\begin{table}[!htbp]
    \centering
    \begin{tabular}{llll}
        \toprule
            & Fruit Fly & \blue{KKI113} & European grid \\
        \hline
        $d_g$ & 5.4(1)~\cite{Flycikk} & \blue{3.4(1)~\cite{sniccikk}} & 2.6(1)~\cite{USAEUPowcikk} \\
        $d_s$ (unwei.)  & 3.375 & \blue{--}   & 1.717 \\
        $d_s$ (wei.) & 2.342 & \blue{--} & 1.507  \\
        \hline
    \end{tabular}
    \caption{The graph dimensions and spectral dimensions of the FF connectome and the European grid network. Both spectral dimensions for unweighted and weighted networks are given. \blue{Spectral dimensions of the KKI113 graph were not computed in the lack of sufficient computer resources.}}
    \label{tab:ds}
\end{table}

To compute the spectra dimensions, we extracted the cumulative density distributions of the first 200 eigenvalues of both the Laplacian matrices of the unweighted and weighted FF connectomes and plot them in a log-log scale as shown in Fig.~\ref{fig:dsff}. For small enough $\lambda$ values, the distributions indeed display a scaling regime,  permitting estimation of spectral dimensions by Eq.~\eqref{eqs:dsc}, as listed in the plot legend and Table~\ref{tab:ds}. Now, even though $d_g^{FF}>4$, since $d_s<4$, one should expect frustrated synchronization in some parameter regime where \red{chimera-like} states may be observed.

\begin{figure*}[!htbp]
\begin{center}
\includegraphics[width=0.99\textwidth]{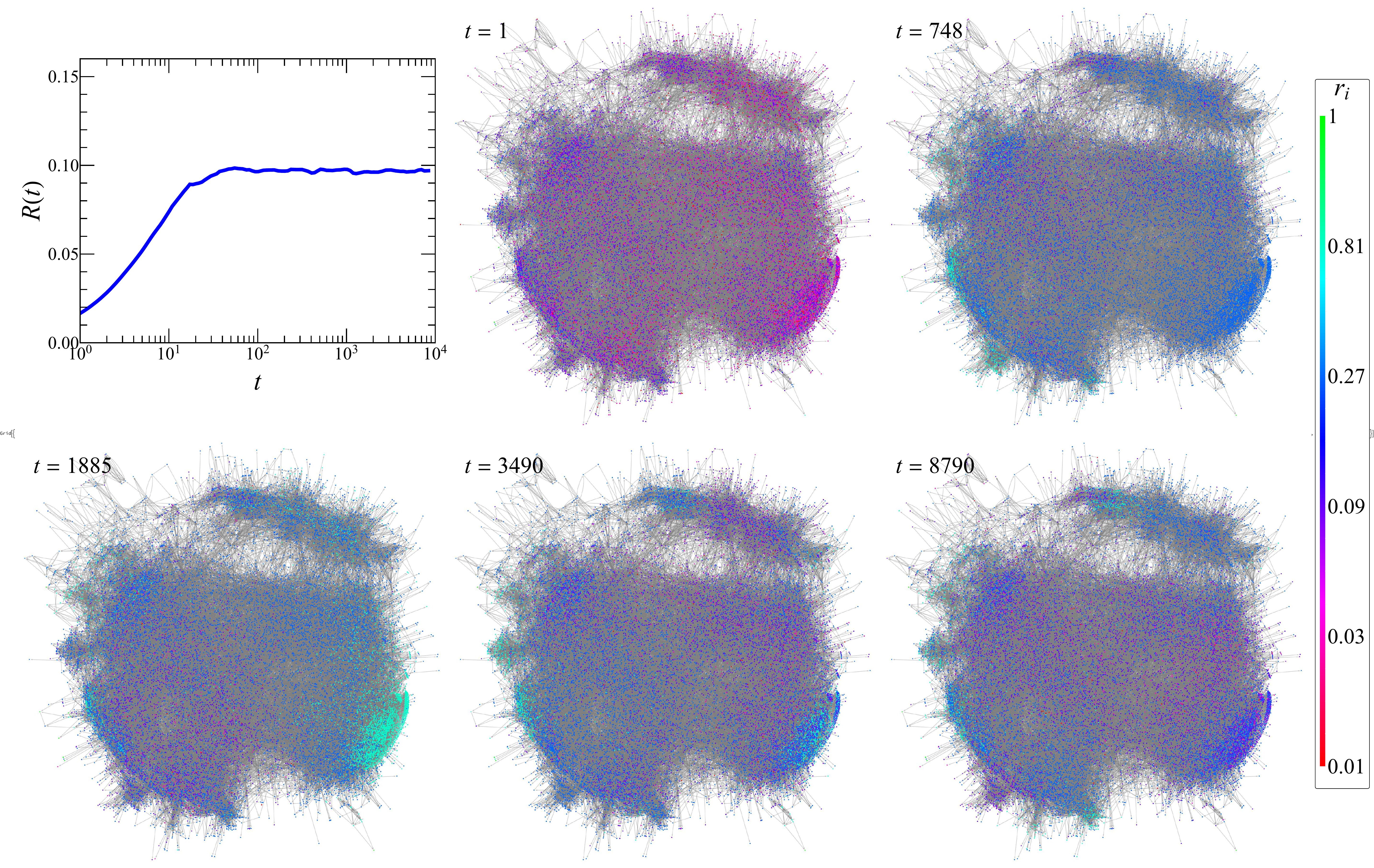}
\caption{\label{fig:ff} Evolution of $R(t)$ towards the steady state following a start from a fully unsynchronized state (upper left panel) and local Kuramoto results at five different time steps encoded by the color map of $r_i$ for the weighted FF connectome. Results were averaged over 20 samples. In the colormaps, red corresponds to low local synchronization, and green to high synchronization.}
\end{center}
\end{figure*}

To that end, we solved Eq.~\eqref{kureq} and Eq.~\eqref{diffeq} numerically with respect to different coupling strengths and $F$ on the weighted FF connectome and observed the respective global phase order parameter $R(t\to\infty)$ in the stationary state. The stationary state is typically reached after a few hundred time steps; see, for example, the first panel of Fig.~\ref{fig:ff}. Practically, we followed the dynamics up to $t=1000$ to ensure stationarity. In Ref.~\cite{Flycikk}, we had estimated the critical coupling $K_c\simeq 1.6$ from the peak of the variance of $R(t\to \infty)$ with respect to $K$. By using this critical coupling, we further calculated the local order parameter Eq.~\eqref{LCO} for each node, averaged over 20 independent simulation runs. In Fig.~\ref{fig:ff}, the local order parameters for three representative time steps are displayed by encoding the respective values to a color map. Since the simulations were started from a fully asynchronous state, we see that the system gradually evolves into a more synchronous state at larger times. However, even in the globally stationary state characterized by a constant $R$ value, the local order parameters show rather inhomogeneous patterns, with some parts of the connectome are more synchronized (greener regions) while some other parts are less synchronized (redder regions), indicating the emergence of \red{chimera-like} states~\cite{restrepo2005,schroder2017}. Note the disparity of the synchronization levels between different regions is quite large in this case, with greener regions almost fully synchronized and red regions fully unsynchronized.  What is more, as partially shown by the second and the third panels at $t=748$ and $t=1885$, the distribution of the local order parameters can still evolve in the globally stationary state. Simulation results seemed to suggest quite random temporal behavior for the local order parameters (not shown here), but more careful studies for the long-time behavior are still needed to examine if it is periodic with a very long period. These results are thus suggestive of strong spatiotemporal fluctuations in \red{chimera-like} states, as it is typical for frustrated synchronization~\cite{millan2018}.

\begin{figure}[!ht]
    \centering
    \includegraphics[width=\columnwidth ]{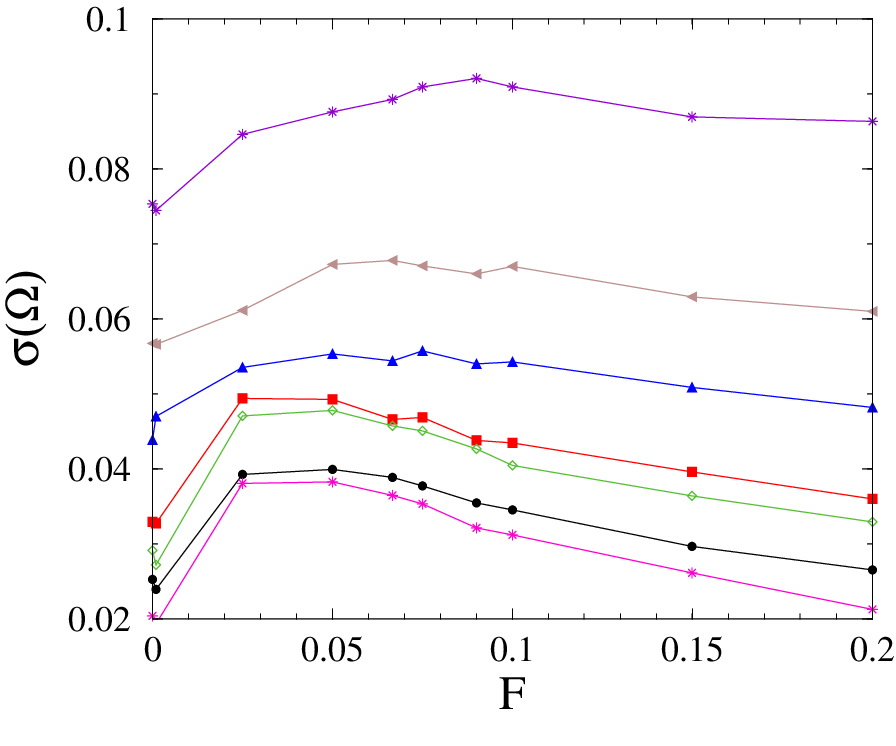}
    \caption{Community dependence of $\sigma(\Omega)$ at $K=1.6$ in the FF. Peaks, marking the
    local entrainment point $F_c'$, move from 0.025 to 0.1 for communities 8,6,4,3,2,1 and the
     whole (top to bottom curves). Results were obtained by averaging over 200 samples.}
    \label{fig:R-F-9}
\end{figure}

To provide more evidence of the \red{chimera-like} states, we have calculated the order parameters in the steady state
at $K=1.6$ in the nine largest communities, determined by the Louvain method~\cite{sniccikk}.
However, the community dependence is rather weak in case of the Kuramoto model. So we enhanced the
local synchronization by adding periodic forces within the framework of the SK model.
The transition point shifts in the range $F_c \in [ 0.05, 0.1 ]$. 
Even more evident community dependence could be found in the frequency synchronization points estimated by the peaks of the variances of the order parameter $\Omega$.
As one can see in Fig.~\ref{fig:R-F-9}, frequency entrainment occurs in the range $F'_c \in [ 0.025, 0.1 ]$ in different communities. That means that for certain forces some communities are in the super-, while others are in the sub-critical states locally, suggesting \red{chimera-like} states.

As it has already been shown in Refs.~\cite{Flycikk,sniccikk} the FF graph exhibits a weak
modular structure. Much higher level of modularity can be observed in human connectomes, albeit on a coarse grained scale, describing the white matter. The large human connectomes obtained by Diffusion Tensor Imaging of MRI~\cite{MIG,CCcikk,Flycikk,sniccikk} has node numbers of order of a million. We have not been able to
calculate the local order parameters and the spectral dimensions for such large systems. 
In \cite{CCcikk} we calculated their graph dimensions, which proved to be above 3 of the 
embedding space, but lower than 4, due to the long fiber tracts, connecting distant regions. 
We show here, that synchronization of communities of the KKI-113 graph exhibits much more visible 
differences than that of the FF, suggesting strong \red{chimera-like states} in case of the Kuramoto model running on them. This small-world graph contains $799\,133$ nodes, connected via $48\,096\,500$ 
undirected (\blue{symmetric)} and weighted edges and exhibit a hierarchical modular structure, 
because it was constructed from cerebral regions of the Desikan--Killany--Tourville parcellations, which is standard in neuroimaging~\cite{DESIKAN2006968}.
\blue{These parameters come from the database after discarding the unconnected parts.}
The modularity quotient is much higher, than that of the FF: $Q_{KKI-113} \approx 0.915$ 
and the topological dimension is just $d_g=3.4(1)$~\cite{sniccikk}. 

\begin{figure}[!htbp]
    \centering
    \includegraphics[width=\columnwidth ]{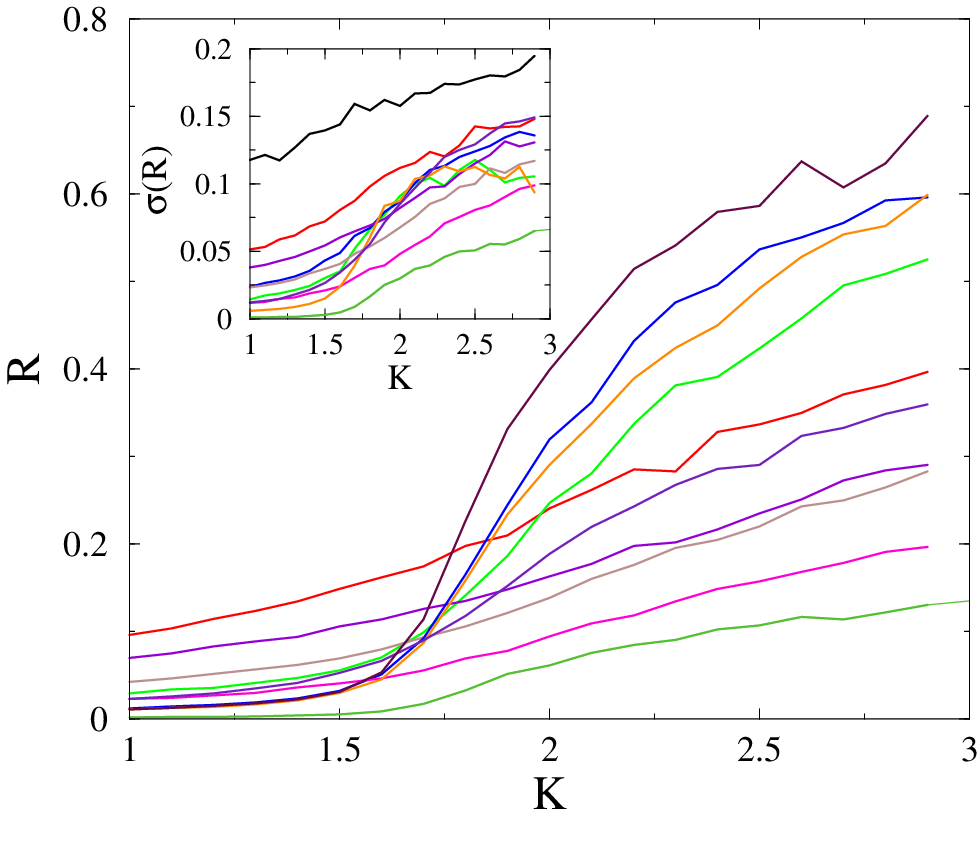}
    \caption{Community dependence of $R$ for different $K$-s showing different
    phase synchronizations, corresponding to \red{chimera-like} states in the KKI-113 connectome. 
    The the lowest curve denotes the synchronization of the whole system, which grows the
    slowest by increasing $K$.
    Inset: Fluctuations of the same data, showing peaks at different $K$-s.
    The lowest curve, representing the whole system grows the slowest. 
    Results were averaged over 100 samples.
    }
    \label{fig:R_K_113}
\end{figure}

\blue{
As one can see on Fig.~\ref{fig:R_K_113} in certain communities the phase synchronization is high for $K > 1.7$, while others are still practically unsynchronized at this coupling. 
Similarly, for $K < 2.2$ the frequency order parameter $\Omega$ is close to zero for 3 communities and $\Omega > 0.4$ for the others as shown in the Appendix. These suggest \red{chimera-like} states in the $1.7 < K < 2.2 \ $ region.}

\subsection{\red{Chimera-like} states in the European high-voltage power grid network \label{sec:3A}}

Unlike neural networks on which the oscillators are massless\blue{, power-grid networks } are massive and should be described by the second-order Kuramoto model. In this subsection, we attempt to show if \red{chimera-like} states can emerge in such systems. Power-grid networks are genuinely hierarchical modular networks if the detailed information for the medium- and low-voltage parts of the grids are also incorporated. Practically, it is almost impossible to infer the entire structure of large power-grid networks, but it is feasible to mimic it
by adding medium- and low-voltage parts to the high-voltage (HV) skeleton, according to 
the empirical hierarchical distribution, as it was done in Ref.~\cite{POWcikk}
We downloaded the European HV power grid from the ``SciGRID Dataset''~\cite{EUgrid} encoding the 2016 status, deduced via processing google street-map. We have not supplemented this graph with lower-voltage parts, but it already contains $12$ kV, $20$ kV, ... links, which belong to the middle-voltage category, according to the definition of the $100$ kV threshold for HV lines.

This graph contains $N=13478$ nodes, interconnected via $E=33844$ links.
\blue{This database contains weightless connections between nodes. We assumed undirected edges, as theoretically power can flow in both direction on HV links. After symmetrizing the edge weights, i.e. $W_{ij}=W_{ji}$} an average degree $\langle k\rangle =2.51$ was obtained.
In Fig.~1 of Ref.~\cite{USAEUPowcikk} the degree distribution is shown. 
The tail of the degree distribution for $k\ge 15$ could be well fitted by a 
stretched exponential $8.25 \times e^{-0.53(5)k}$ function, which renders this network at the threshold of robust/fragile: $\gamma=3/2$, as according to the definition in~\cite{Pgridtop}, networks with a $P(k>K)=C e^{-k/\gamma}$ cumulative degree distribution and $\gamma < 3/2$ are robust, based on a mean-field percolation theory under random node removals. The adjacency matrix, visualized on Fig.~2 of Ref.~\cite{USAEUPowcikk} proves that this is a highly modular graph, characterized by $Q^{EU}=0.963$.
Furthermore, it is a small-world network according to the definition of the small-worldness
coefficient~\cite{HumphriesGurney08,USAEUPowcikk}. By calculating the graph dimension
using the breadth-first search algorithm as shown in the inset of Fig.~3 of Ref.~\cite{USAEUPowcikk}, $d_g^{EU}=2.6(1)$ was obtained. 

Since the coupling between a pair of nodes of a power grid is proportional to the maximal power $P_{ij}$ transmitted between them and inverse to the imaginary part $X_{ij}$ of the impedance of the transmission line, weights computed from the normalized values \blue{ $W_{ij} \propto P_{ij}/X_{ij}$~\cite{Heterogeneity} had also been considered to construct a weighted European HV network and analayzed here.}

\begin{figure}[!htbp]
    \centering
    \includegraphics[width=0.9\columnwidth]{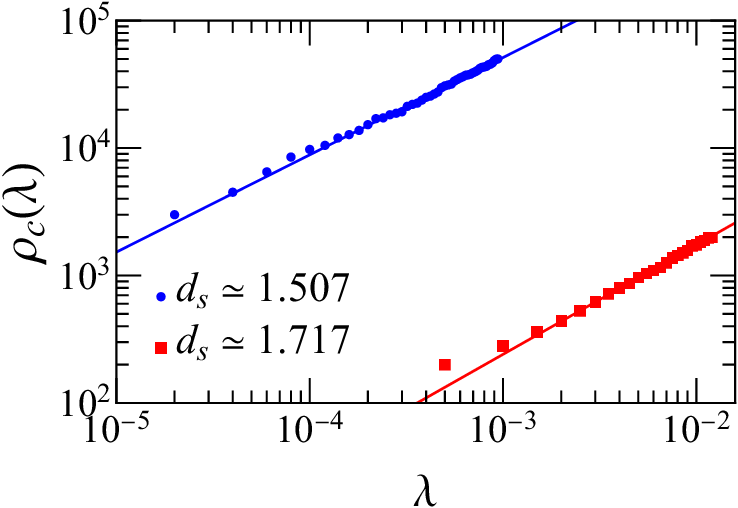}
    \caption{The cumulative eigenvalue densities of the unweighted (red solid square) and the weighted (blue dot) European power-grid network, extracted from the first 200 smallest eigenvalues.}
    \label{fig:dseu}
\end{figure}

By again extracting the cumulative density distributions of the first 200 eigenvalues of both the Laplacian matrices of the unweighted and weighted European power-grid networks, Fig.~\ref{fig:dseu} shows quite clean power laws. As listed in the plot legend and Table~\ref{tab:ds}, the estimated spectral dimensions are both well below the critical dimension $d_c=4$. Hence, one can again expect to observe \red{chimera-like} states, although instead, the second-order Kuramoto model \eqref{kur2eq} is going to be inspected in this case.

\begin{figure*}[!htbp]
\begin{center}
\begin{tabular}{cc}
\includegraphics[width=0.49\textwidth]{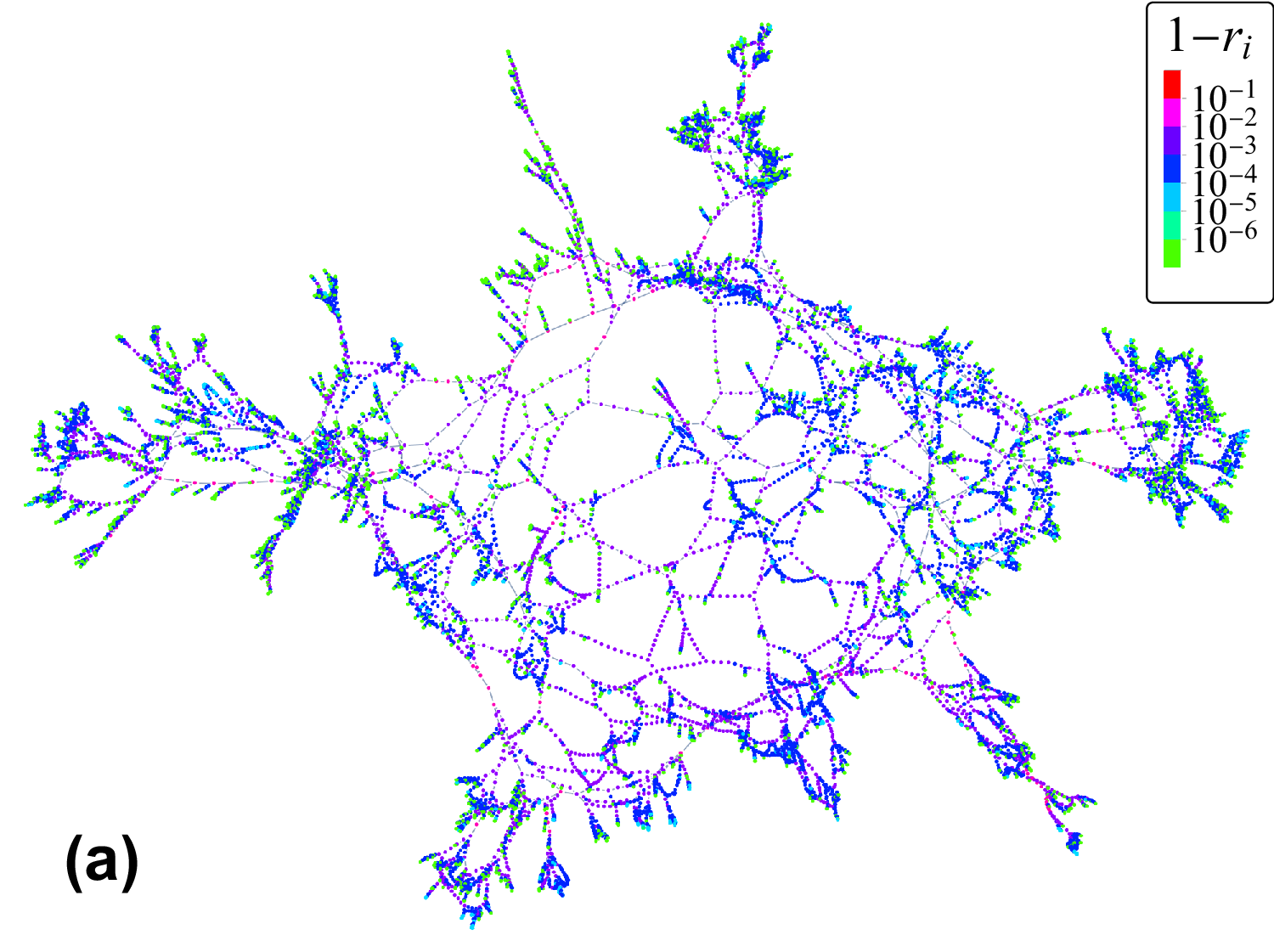} &
\includegraphics[width=0.49\textwidth]{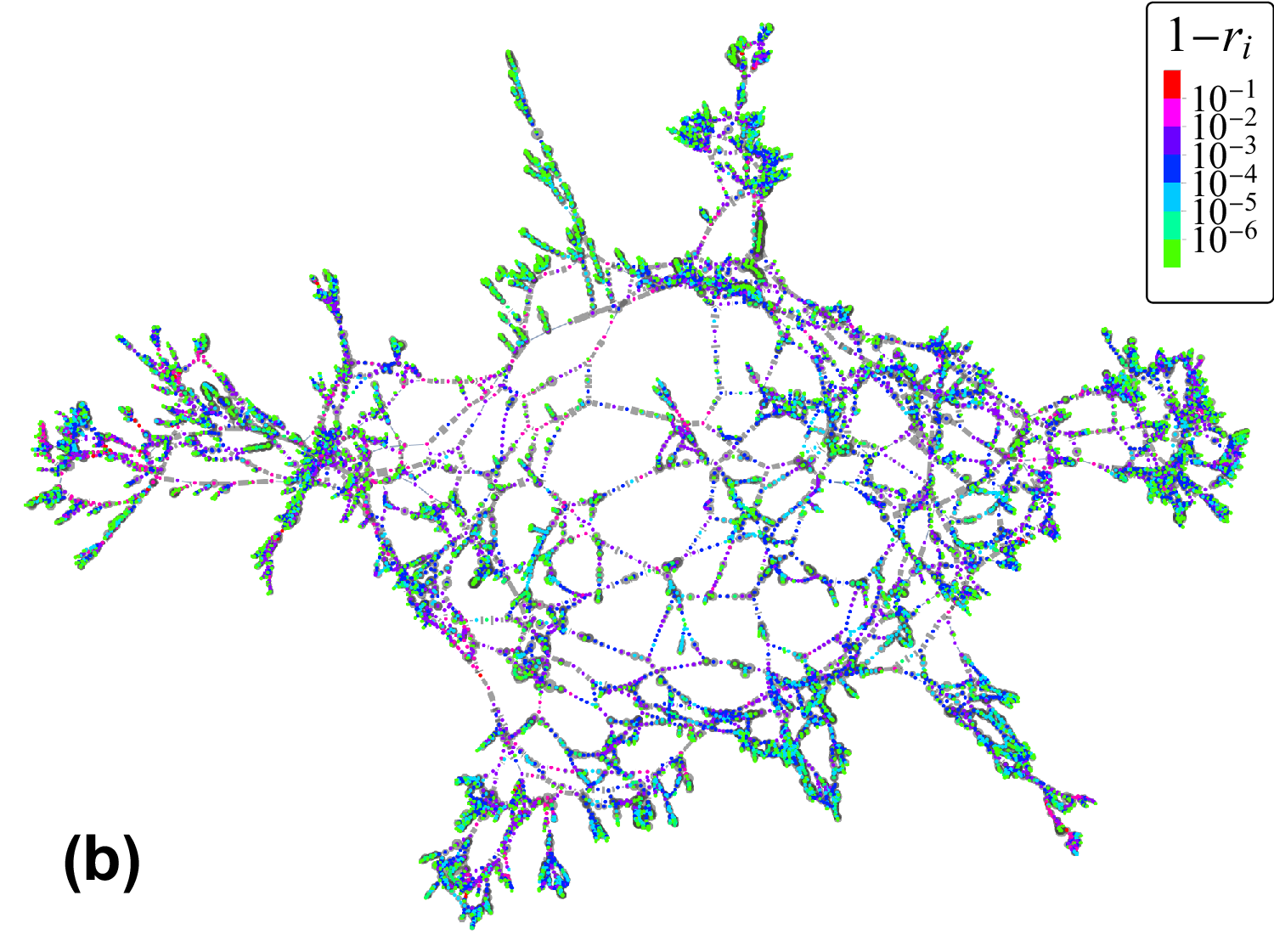}
\end{tabular}
\caption{\label{fig:euch} Local Kuramoto results in the stationary state encoded by the color
maps of $1-r_i$ for (a) the unweighted and (b) the weighted European high-voltage networks. Results were averaged over 20 samples. Red corresponds to low local synchronization, and green to high synchronization. The width of the gray edges in (b) is proportional to the logarithm of the weights.}
\end{center}
\end{figure*}

We again tune the system to the verge of criticality. By solving Eq.~\eqref{kur2eq} to obtain the peak of the variance of $R(t\to\infty)$, the critical couplings had been estimated to $K_c=80$ for the unweighted network~\cite{USAEUPowcikk} and $K_c=7000$ for the weighted network. The stationary states are typically reached after a few hundred time steps (see Ref.~\cite{USAEUPowcikk}), but we solved them up to $t=20000$ to ensure the stationarity of the system.  The local order parameters, calculated in the stationary states after averaging over 100 samples, are then obtained with respect to these critical couplings. Fig.~\ref{fig:euch} shows that inhomogeneous patterns, encoded again in a color map, indeed overwhelm the system. Due to the higher levels of synchronization, there are quite some proportions of oscillators with their local order parameters relatively closer to 1 as compared to the FF connectome case. Hence the color map shows the quantity $1-r_i$ instead. Since the differences in the local parameters in greener regions and redder regions are still quite apparent, we see that as suggested by the low spectral dimension, \red{chimera-like} states indeed can be observed in this case. Note that even though the weighted network is a bit less synchronous globally at $K=7000$ [$R(t\to\infty)\simeq 0.47 $] than the unweighted network at $K=80$ [$R(t\to\infty)\simeq 0.48$], the weighted network still seems to be more synchronized locally in many parts of the network. This emphasizes the importance of incorporating edge weights to take into account more realistic couplings between the nodes.

Comparing the local order parameter patterns in Fig.~\ref{fig:ff} and Fig.~\ref{fig:euch}, it is also interesting to note that less synchronous regions are typically also less clustered as compared to regions with higher levels of synchronization. This is in some sense in reminiscence of the analysis in Ref.~\cite{Villegas_2014}, in which it had been shown that \red{chimera-like} states can also be characterized by the order parameters of different moduli.

\begin{figure}[!htbp]
    \centering
    \includegraphics[width=\columnwidth ]{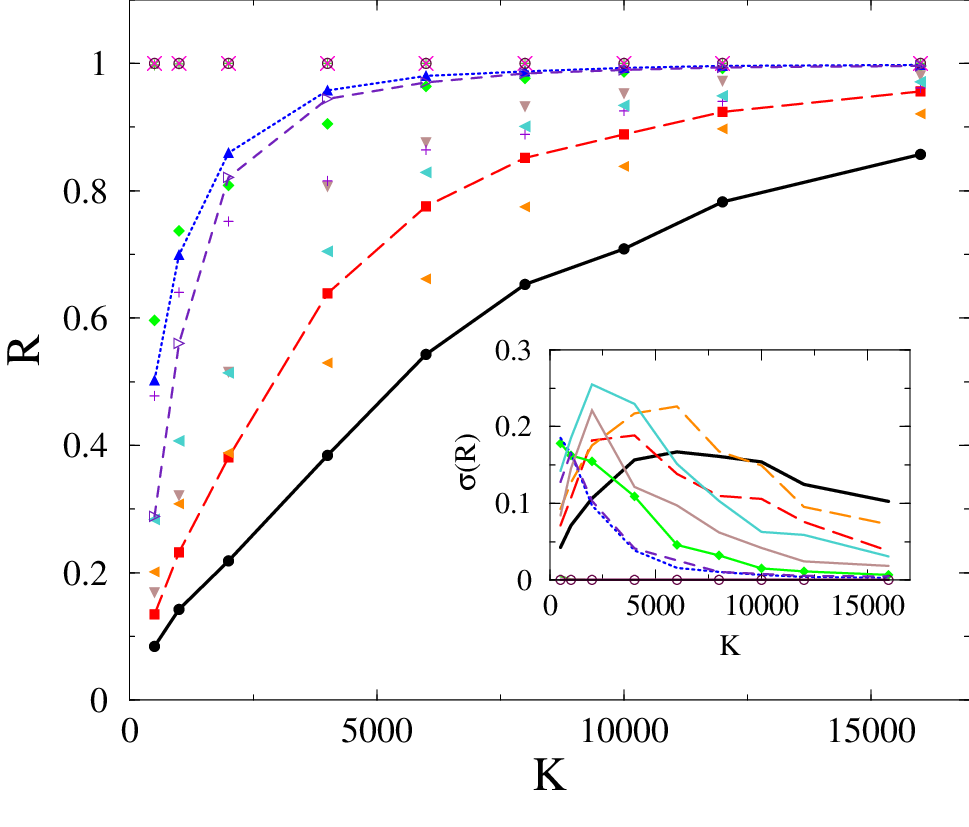}
    \caption{Community dependence of $R$ in the power-grid for different $K$-s showing different
    phase synchronizations, corresponding to \red{chimera-like} states. The thick black
    curve denotes the synchronization of the whole system, which grows the
    slowest by increasing $K$.
    Inset: Fluctuations of the same data, showing different synchronization points.
    The thick black curve, representing the whole system has the rightmost peak.
    Results were obtained by averaging over 100 samples.}
    \label{fig:R-K-12}
\end{figure}

To provide more evidence of the \red{chimera-like} states, we have also calculated the steady-state $R$ in the twelve largest communities, determined via the Louvain method with a modularity score close to the maximum  $Q \approx 0.795$, in the same way as in case of the FF~\cite{Heterogeneity}.
As one can see in Fig.\ref{fig:R-K-12}, synchronization occurs at different couplings
in different communities, such that for small $K$-s the small communities are fully ordered, 
while the larger ones are still desynchronized. This is related to the size dependence of 
$K_c$ in case of crossover, however here the communities are not independent. Note, that the fully ordered communities have less than 100 nodes.

\section{Summary \label{sec:4}}

In this paper, we have demonstrated that \red{chimera-like} states can occur in Kuramoto-type models on large networks if the spectral dimension is low, i.e. $d_s < 4$, even if the graph dimension is not necessarily like that. 
That happened in case of the graph of the FF connectome, which exhibits $d_g = 5.4(1)$. This is in agreement with the hypothesis, advanced for the first-order Kuramoto model in~\cite{millan2019}. But as modularity is weak for FF, so do the \red{chimera-like states}. We can show them by a community-level analysis with an applied periodic external field. In contrast, for a large human connectome, possessing high modularity, we show strong community dependence of the local synchronization.
We used the novel, more general definition of ``\red{chimera-like} states'', introduced for single-component models on complex networks~\cite{Scholl2016}. According to this spatially bounded regions of partial synchronization emerge in extended control parameter spaces, similarly as in case of GP-s. This is different from the phase coexistences in multi-component systems or the spinodal decomposition at a first order phase transition point.

Power grids can be described by the second-order Kuramoto model, which possesses inertia.
We found that the European HV power grid has a graph dimension $d_g = 2.6(1)$, 
but the spectral dimensions seem to be below $d_s=2$. Still, the occurrence of \red{chimera-like}-like patterns can be observed via order parameters and confirmed by a community-level synchronization study.
We demonstrated the level of local synchronization by showing the local Kuramoto order parameter, but similar results have been found by calculating the local frequency spreads.

\begin{acknowledgments}
We thank Krist\'of Benedek and B\'alint Hartmann for providing weight calculations of the European network, Istv\'an Papp for exploring the communities, Jeffrey Kelling for developing the GPU solver code, and R\'obert Juh\'asz for the helpful discussions.
This research was funded by ELKH grant SA-44/2021, and the Hungarian National Research, Development, and Innovation Office NKFIH grant K128989, \red{K146736}. Most of the numerical work was done 
on KIF\"U supercomputers of Hungary.
\end{acknowledgments}

\section*{Appendix}

\begin{figure}[!htbp]
    \centering
    \includegraphics[width=\columnwidth ]{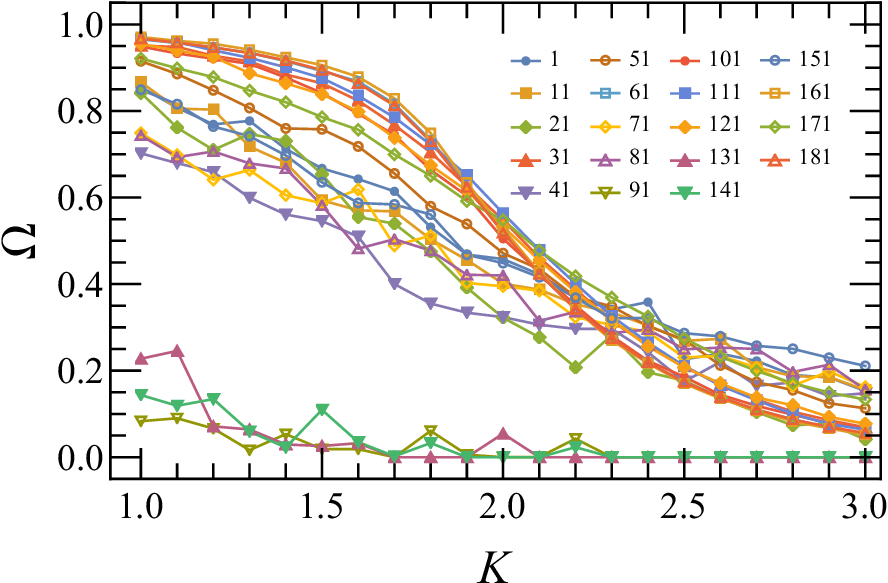}
    \caption{Frequency entrainment of the Kuramoto model on KKI113  
    in communities, denoted by the legends for different $K$-s. The steady state $\Omega$ order parameter shows different levels of
    synchronizations, corresponding to \red{chimera-like} states.
    Results were obtained by averaging over 100 samples.}
    \label{fig:O-K-12}
\end{figure}

Here we show community and $K$ dependence of $\Omega$ of the Kuramoto model on the KKI113 connecotme. As one can see \red{on Fig.~\ref{fig:O-K-12}} for the weakly coupled, $K < 2.2$ region most of the system remains frequency unsynchronized, while communities \mage{91, 131, and 141} exhibit frequency entrainment corresponding to a \red{chimera-like} state.

\section*{Data Availability Statement}

Data are available on request from the corresponding author.

\bibliography{aipsamp}% Produces the bibliography via BibTeX.

\end{document}